\documentclass{nseJournal}

\usepackage{amsmath}%
\usepackage{amsfonts}%
\usepackage{amssymb}%
\usepackage{amstext}
\usepackage[colorlinks=true,allcolors=blue]{hyperref}
\usepackage{comment}
\usepackage{orcidlink}
\usepackage{upgreek}
\usepackage{cellspace} %
\begin{document}

\title{Ion Temperature Inference from Neutron Counting in Maxwellian Deuterium Plasmas}

\addAuthor{Allison J.\ Radich}{a} 
\correspondingEmail{pat.carle@generalfusion.com}
\addAuthor{Vlad Grecu}{a}


\addAuthor{\correspondingAuthor{Patrick J. F. Carle}}{a} 

\addAuthor{Myles {Hildebrand}}{a}

\addAuthor{Stephen J.\ Howard}{a}

\addAuthor{Colin P.\ McNally}{a} 

\addAuthor{Meritt Reynolds}{a}

\addAuthor{Akbar Rohollahi}{a}

\addAuthor{Ryan E. Underwood}{a} 

\addAuthor{Sara Weinstein}{a} 

\addAffiliation{a}{General Fusion Inc.\\ 6020 Russ Baker Way, Richmond, British Columbia V7B 1B4, Canada}

\addKeyword{plasma diagnostics}
\addKeyword{ion temperature}
\addKeyword{neutron detection}
\addKeyword{scintillator}
\addKeyword{pulse shape discrimination}

\titlePage 

\begin{abstract}

A method is presented for inferring the deuterium fuel ion temperature from neutron counts measured with fast liquid scintillators in conditions where the ion velocity distribution is Maxwellian. 
Local neutron count rates at each scintillator position are combined to estimate total neutron yield from the plasma, where absolute detection efficiency is determined via MCNP neutron scattering simulation based on a 3D model of the experiment structure. 
This method is particularly advantageous for Magnetized Target Fusion applications as it yields a time-resolved diagnostic and does not require direct line-of-sight to the plasma or collimation of the neutrons.
The instrumentation configuration, pulse-shape discrimination and pile-up correction algorithms, detector calibration, and ion temperature calculation method with uncertainty characterization are discussed.
An application of the method to General Fusion's Plasma Injector~3 (PI3) spherical tokamak device is demonstrated and the results are compared to an Ion Doppler spectroscopy ion temperature diagnostic.
\end{abstract}

\section{Introduction}\label{intro}

Ion temperature ($T_i$) is an essential plasma parameter in fusion science, required to understand the heating and transport properties of plasma and thus to control plasma performance and fusion efficiency. Typical approaches to determining $T_i$ rely on direct line-of-sight access to the plasma, for example, using charge-exchange or Ion Doppler spectroscopy, a time-of-flight neutron spectrometer, or laser-based systems such as Thomson scattering \cite{knoll_radiation_2010, isler_overview_1994, wolle_tokamak_1999, firk_neutron_1979, elevant_jet_1991}. 

Diagnostic access to the plasma in Magnetized Target Fusion (MTF) experiments can be challenging, due to the collapse of the compressed liner around the plasma. The liner occludes views into the plasma, with unobstructed sight lines for conventional spectroscopic or laser-based diagnostics available only through the ends of the fusion vessel \cite{carle_neutron_2022}. Further, in the later stages of compression, it is likely that all access to the plasma will be blocked. There is therefore a strong interest in methods of estimating $T_i$ that do not rely on direct access to the plasma.

In the absence of direct neutron energy measurements, a common approach is inference of $T_i$ from neutron yield assuming a Maxwellian distribution for ion velocity, which can be measured in several different ways \cite{artsimovich_investigation_1972, ekdahl_1979, wolle_time-dependent_1994}. One possibility is neutron activation analysis which uses materials that become radioactive when exposed to neutrons. After exposure, the decay products of the activated material, typically beta or gamma rays, are measured to determine the neutron yield \cite{zankl_neutron_1981, strachan_neutron_1990, jarvis_use_1991}. With sufficient neutron yield, activation analysis can provide a total neutron count from a plasma discharge, but not a time-resolved neutron rate, although the total count can be used as corroborating evidence for measurement of $T_i$. 

In order to obtain a time-resolved rate during a plasma discharge, it is necessary to use neutron counters. Gas proportional counters exploit the ionizing interaction between the emitted neutrons and the gas atoms (e.g., Helium-3), which creates a charge that can be measured as an electrical pulse indicating a neutron event \cite{batchelor_helium3_1955, jarvis_neutron_1994}. Gas proportional counters have the advantages of being relatively insensitive to gamma radiation and sensitive to neutrons, providing accurate neutron counts for relatively low neutron yield.
However, their low time resolution cannot measure the evolution of the neutron rate over the relatively short span of a milliseconds-long MTF pulse.

 \begin{figure*}
     \centering
     \includegraphics[width=0.7\textwidth]{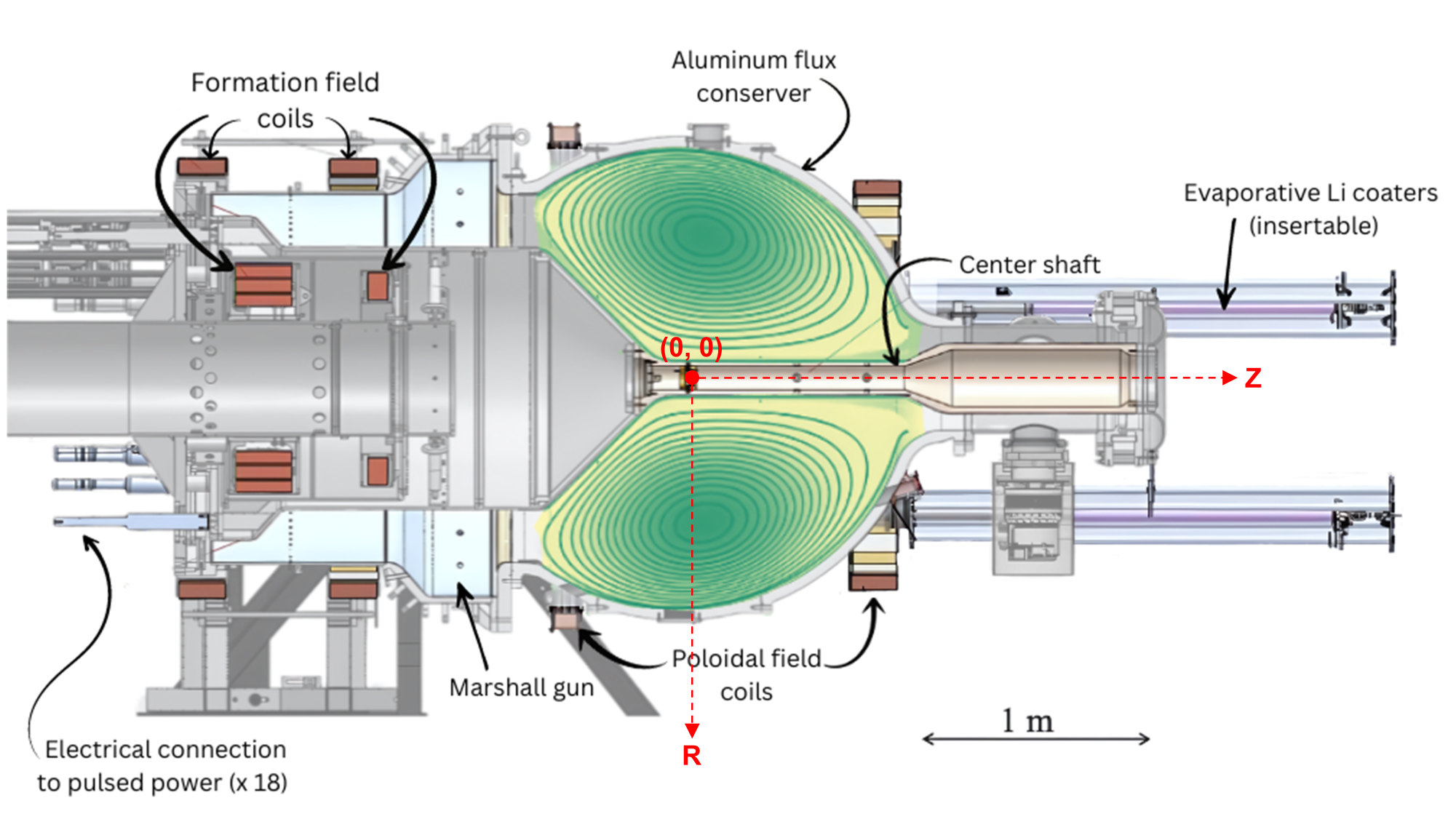}
     \caption{Diagram of PI3 with the origin of the coordinate system shown at the center of the flux conserver. The Marshall gun forms the initial plasma and pushes it into the flux conserver \cite{tancetti_thermal_2025}.}
     \label{fig:pi3-coord}
 \end{figure*}

For relatively short-lived MTF plasmas, with typical lifetimes ranging from a few~ms to a few~$\upmu \text{s}$, neutron detectors with a fast response such as scintillators (decay time $< 0.2\,\upmu\text{s}$) are needed to obtain time-resolved neutron counts. Scintillators are luminescent materials that emit light when excited by high-energy radiation including neutrons and electromagnetic radiation (notably gamma rays). Scintillator-based neutron detectors are composed of a scintillating material coupled with a photodetector, commonly a photomultiplier tube (PMT) or a photodiode, which converts the light emitted from the scintillator to an electrical signal \cite{knoll_radiation_2010, birks_theory_1964}. For the purposes of neutron detection in MTF experiments with relatively low neutron emission and rapidly evolving plasmas, organic liquid scintillators provide a robust way to detect neutrons in a mixed neutron and gamma-ray field, since the two types of interactions emit light with distinguishable decay time-constants \cite{batchelor_response_1961}. Neutrons interact primarily with the protons of the scintillator molecules via elastic scattering, whereas gamma rays interact with the atomic electrons of the scintillator atoms generating Compton-scattered electrons. These electrons decelerate more rapidly than an energetic proton from a neutron impact. The longer tail of light emission from a proton slowing indicates that an event was a neutron rather than a gamma ray. The process of distinguishing between neutron and gamma-ray events based on this difference is known as pulse shape discrimination (PSD) \cite{batchelor_response_1961, ROUSH1964112}.

This paper presents a method for inferring core $T_i$ using neutron detection with uncollimated fast liquid scintillators, building on an approach originally developed for a series of magnetized plasma compression experiments known as the Plasma Compression Science (PCS) program \cite{Howard_PCS16}. The method has been extended to estimate $T_i$  from a spherical tokamak device called Plasma Injector~3 (PI3) operated from 2017 to 2024 (Fig.~\ref{fig:pi3-coord}) \cite{tancetti_thermal_2025}.

The design of PI3 used a coaxial Marshall gun to form a spherical tokamak plasma within an aluminum flux conserver. The plasma was confined by a poloidal magnetic field sourced from internal toroidal plasma currents, and stabilized with a toroidal magnetic field generated by axial current driven down the center shaft of the machine. 
Under optimal conditions, PI3 formed plasmas that persisted without external energy input for up to $30\mathrm{\,ms}$ in duration, with core electron temperatures exceeding $400\mathrm{\,eV}$, core electron density $3\times10^{19} \mathrm{\,m^{-3}}$, and emitted up to $10^8$ total neutrons per shot \cite{PI3_APS2024}.
Most PI3 experimental plasmas were primarily composed of deuterium, with an ion temperature that was high enough to produce a detectable neutron yield from deuterium-deuterium (D-D) fusion events.  The neutrons produced with an energy of $2.45\mathrm{\,MeV}$ traveling outward through the vessel provided a diagnostic opportunity to infer the core ion temperature based on known D-D fusion physics. This paper describes the instrumentation setup used to measure neutron emission, and a method to infer $T_i$ from neutron counts in combination with measurements of the plasma density profile.

\section{Instrumentation Setup}\label{inst-setup}
The detector array used four uncollimated liquid scintillators manufactured by SCIONIX Holland B.V.\ model numbers: VS-1105-21 (2, 0.83 L, Eljen EJ-309), VS-1025-13 (1, 3.49 L, Eljen EJ-301), VS-1025-14 (1, 3.49 L, Eljen EJ-301) \cite{Eljen}. Each had a cubic can containing scintillating fluid coupled to a PMT from ET Enterprises Ltd. (9821B series 78-mm diameter or 9390B series 130-mm diameter). The scintillators were enclosed in steel boxes lined with 25.4 mm lead shielding to attenuate the transmission of photons (X-rays and gamma rays) originating from the plasma or from fast particle interactions with surrounding materials. The pulse height response of the scintillators was calibrated with cobalt-60 (Co-60) absolutely calibrated gamma-ray sources at General Fusion.

A data acquisition and control cart contained an 8-channel Keysight MXR058A oscilloscope to record scintillator signals at a sampling rate of 1 GS/s, and high-voltage bias power supplies for the PMTs. Electrical connections between the scintillators and the control cart were made with $50\,\Omega$ impedance coaxial cables enclosed within a grounded copper braided shield line with a star ground configuration. Earth ground was made at the control cart. 
The anode output signal from each PMT was split into two scope channels: high and low gain. The low-gain channel covered the maximum voltage range expected for a signal pulse originating from a $2.45\mathrm{\,MeV}$ neutron. The high-gain channel had a smaller voltage range, and was more sensitive to low-amplitude signals. When high- and low-gain signals were found to be correlated, they were merged to extend the dynamic range of the measurement. 

\begin{figure}
     \centering
     \includegraphics[width=0.5\linewidth]{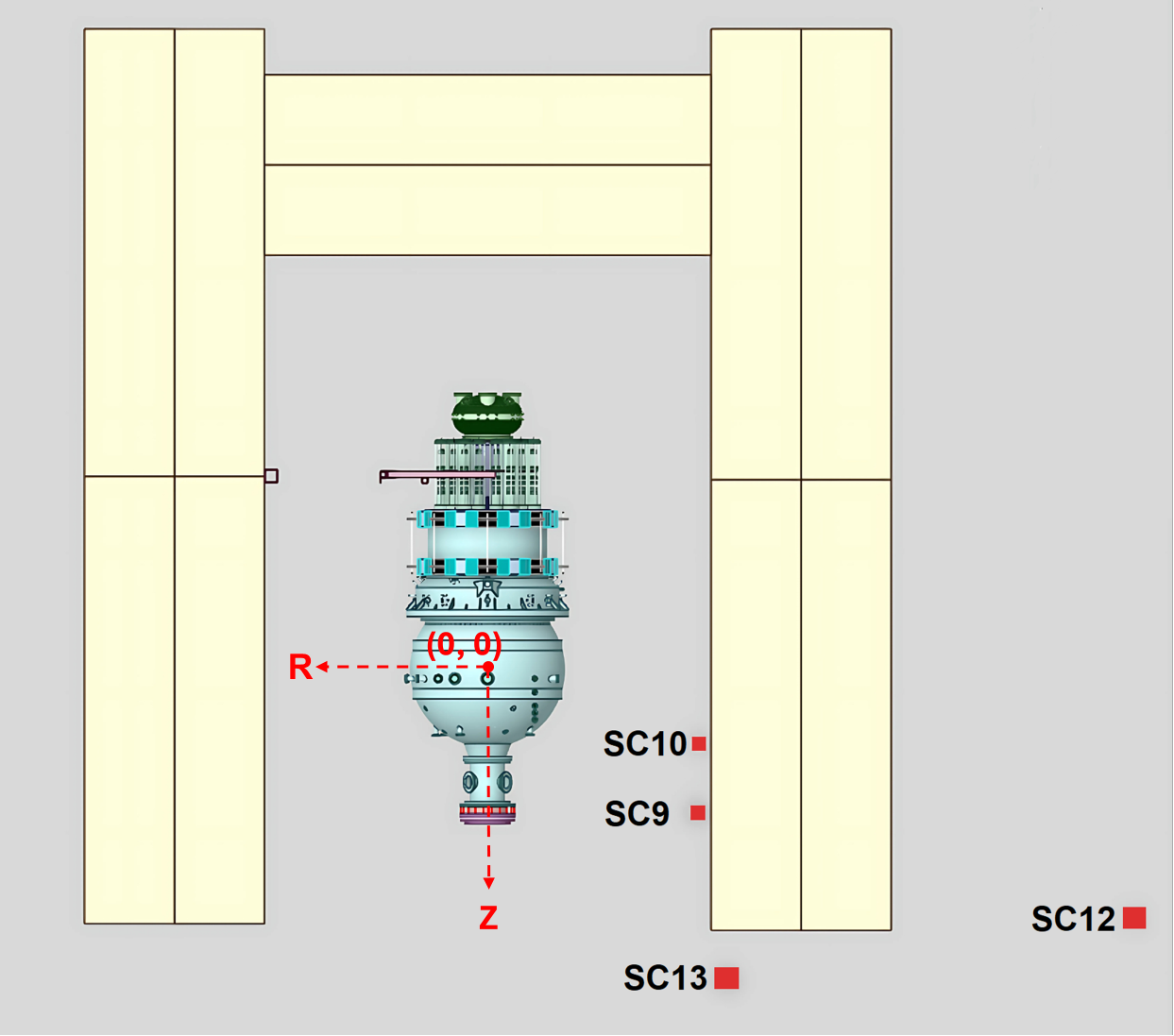}
     \caption{Top view of the PI3 lab space with scintillator locations shown. The origin of the coordinate system is  at the center of the flux conserver. Not shown, $\theta=0$ is directed vertically out of the page.  The yellow boxes represent the containers for the pulsed power system.}
     \label{fig:scint-locs}
 \end{figure}

Table~\ref{tab:scint-sum} provides additional details on the hardware and setup, including scintillator liquid type and volume, location vectors as $(R, Z, \theta)$, absolute neutron detection efficiency and efficiency relative errors as determined by Monte Carlo simulation. The scintillator locations used on PI3 as of May 2024 are tabulated relative to the center of the flux conserver illustrated in Fig.~\ref{fig:scint-locs}. The method of determining the absolute neutron detection efficiencies needed to infer an ion temperature by a Monte-Carlo simulation, taking into account the geometry and materials present, is outlined in Section~\ref{neutron-efficiency}.

\begin{table*} 
    \centering
    \caption{Scintillators in use after May 8, 2024. Absolute neutron detection efficiency $\mathcal{E}(E_\mathrm{thr})$ is for the corresponding proton recoil energy threshold $E_\mathrm{thr}$, calculated from experimental equivalent data. The relative error is the error on the neutron detection efficiency divided by the detection efficiency.}
    \label{tab:scint-sum}
    \begin{tabular}{Sccccccccc}
        \hline
         Scintillator & Fluid type & Volume [L] & $R$ [m] & $Z$ [m] & $\theta$ [deg] & $E_\mathrm{thr}$ [MeV] & $\mathcal{E}(E_\mathrm{thr})$ & Relative error \\
         \hline
         SC9 & EJ-309 &  $0.83$ & $3.33$ & $1.94$ & $124$ & 1.0 & $2.66\times 10^{-5}$ & $0.017$\\
        SC10 & EJ-309 &  $0.83$ & $3.24$ & $1.01$ & $121$ & 1.0 & $3.55\times 10^{-5}$ & $0.019$\\
        SC12 & EJ-301 &  $3.49$ & $8.75$ & $3.37$ &  $99$ &	0.5 & $1.52\times 10^{-5}$ & $0.025$\\
        SC13 & EJ-301 &  $3.49$ & $3.55$ & $4.19$ & $118$ &	0.5 & $6.26\times 10^{-5}$ & $0.009$\\
        \hline
    \end{tabular}
\end{table*}

\section{Temperature Calculation Inputs}\label{calc-inputs}
There are three inputs required to calculate ion temperature using the method described in this work: (1) the neutron emission rate (yield) determined by the liquid scintillator neutron counting; (2) the plasma density profile as a function of poloidal flux coordinate, where density is assumed to be a flux function; and (3) the volume per shell of the plasma’s flux surfaces. 

The axisymmetric density distribution in the plasma volume is an output of a magnetic reconstruction code \cite{tancetti_thermal_2025, BayesianRecon_APS2022}. The main diagnostic inputs for reconstruction include the poloidal and toroidal magnetic fields measured by Mirnov coils located on the wall of the machine, and the interferometer line average density measured from chords spanning the plasma from edge to core. The following sections will provide details of how these ion temperature calculation inputs are obtained, how uncertainties are estimated, and any assumptions made in the process.

\subsection{Neutron Counting with Pulse Shape Discrimination}\label{neutron-count}

For each signal recorded by a scintillator with a pulse height above a set threshold, PSD is performed to classify that event as a neutron or gamma ray \cite{ROUSH1964112}. The pulse height threshold is set to be equivalent to a proton recoil energy of $1\mathrm{\,MeV}$ for scintillators SC9 and SC10, and $0.5\mathrm{\,MeV}$ for scintillators SC12 and SC13, where a proton recoil is produced by elastic scattering of an incident neutron. Each scintillator pulse is integrated using predefined integration time windows characteristic to each scintillator to determine the pulse’s total area and tail area. The ratio of these areas is compared to a power-law bifurcation curve to determine whether the pulse originates from a neutron or gamma-ray interaction \cite{knoll_radiation_2010}. The bifurcation line has been calibrated for each scintillator using a Co-60 gamma-ray calibration source by fitting the edge of the PSD distribution using a power law. 

Adjacent pulses on the same scintillator, which have distinct peaks but partially overlapping integration time windows, are excluded from PSD processing because information is lost in the superposition of two unknown waveforms, making the simple tail-to-total ratio no longer effective at distinguishing between neutrons and gamma rays. This situation of two overlapping pulses is referred to as ``pile-up''. A special case of pile-up happens when the inter-arrival time of two detection events is less than the intrinsic pulse width, in which case the peak finding algorithm will not count two pulses, but one larger pulse. This case of ``hidden'' pile-up can not be directly detected, but is only estimated statistically based on the instantaneous rate of the Poisson process and the average width of scintillator pulses. However, it is generally a small effect (less than 10\% of the rate of distinguishable piled-up pulses) for most count rates on PI3 plasmas, and so will be neglected in this work. The hidden pile-up would become around 5\% of the rate in the scintillator if the pulse rate were over $5\times10^6~\rm{counts/s}$ for a prolonged period. The rate of observable pile-up pulses can be quantified and corrected for. The tabulation of pile-up correction rates is performed in a downstream data processing stage.

\begin{figure}
     \centering
     \includegraphics[width=0.5\linewidth]{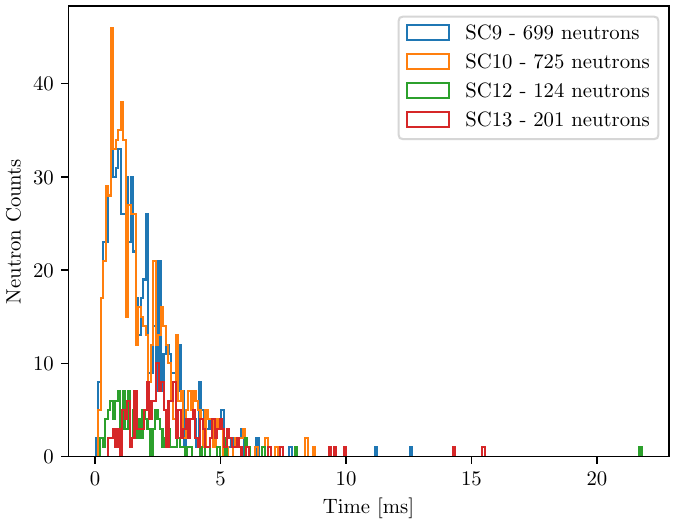}
     \caption{PSD neutron counts as a function of shot time for scintillators SC9, SC10, SC12 and SC13 for shot 22714 (prior to pile-up correction).}
     \label{fig:psd-example}
 \end{figure}

\begin{figure}
     \centering
     \includegraphics[width=0.5\linewidth]{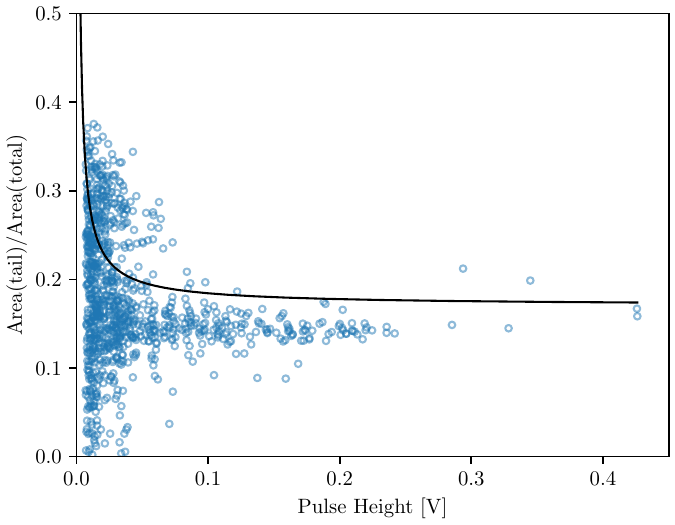}
     \caption{PSD plot with the tail-to-total pulse area as a function of pulse height for SC13, shot 22714. The bifurcation line divides the D-D fusion neutrons (above) from gamma rays (below).}
     \label{fig:psd-plot-1}
 \end{figure}

Fig.~\ref{fig:psd-example} shows the binned PSD neutron counts as a function of time for an example shot. Fig.~\ref{fig:psd-plot-1} shows an example of a PSD plot with the ratio of tail-to-total pulse area as a function of pulse height.

The energy calibration of the pulse heights is performed using a Co-60 calibrated gamma-ray source to first obtain an electron recoil energy to pulse height relationship as eV per digitizer Volts, where recoil electrons are produced by Compton scattering of an incident gamma ray. The proton recoil energy is then determined by using the relationship between the scintillation light output of recoil electrons and recoil protons for the scintillation materials used \cite{cecil_1979}.

\subsection{Scintillator Count Rate Binning}\label{rate-binning}

\begin{table*}
    \centering
    \caption{Defined bin thresholds as a function of the number of remaining neutron event counts.}
    \begin{tabular}{cc}
    \hline
         Remaining neutron events  &  Bin threshold for neutron events \\
         \hline
         $\ge 40$& $30$\\
         $15$ --- $39$& $10$\\
         $10$ --- $14$& $5$\\
         $\le 9$& Bin remaining events together\\
        \hline
    \end{tabular}
    \label{tab:bin-size}
\end{table*}

This section describes the process of converting time-resolved neutron events into an aggregate time-resolved neutron rate. The method follows that described in \cite{Howard_PCS16}. The time-resolved neutron events from all active scintillators are combined to form an aggregate list of neutron events. Any event determined to be a neutron from the PSD algorithm but with a corresponding proton recoil energy greater than 2.45 MeV is removed from the aggregate list of neutron events. The aggregate list was then divided into a series of variable time bins, each containing a fixed number of neutron events. The first time bin starts at plasma formation ($t = 0$). A running tally is initiated to count sequential neutron events from the aggregate list until the tally reaches a threshold value. 
A higher threshold value of the counts per bin will result in larger bin sizes that show the general trend of how the rate changes over time, though it is less sensitive to rapid decreases in count rate. Fewer counts per bin will better track rapid changes in the rate with correspondingly higher uncertainties. Table \ref{tab:bin-size} lists the bin threshold values used based on the number of remaining neutron events in the aggregate list, chosen to best track the evolving neutron rate over a wide range of values while maintaining an acceptable level of uncertainty.

Once the count tally reaches the threshold, the first bin is complete and the tally is reset. The first time bin ends and the second time bin begins just before the time of the next sequential neutron event in the aggregate list. The count then continues until the threshold is met again. To prevent double-counting of a single event into two adjacent bins, it is important that bin start and end times do not fall exactly on a detection event time. This process is repeated for the remaining neutron events in the aggregate list. The bin threshold value is adjusted when there are fewer than 40 events remaining in the list (see Table \ref{tab:bin-size}) to provide more time steps near the end of the plasma’s lifetime, when the frequency of events typically slows significantly.

\begin{figure}
    \centering
    \includegraphics[width=0.5\linewidth]{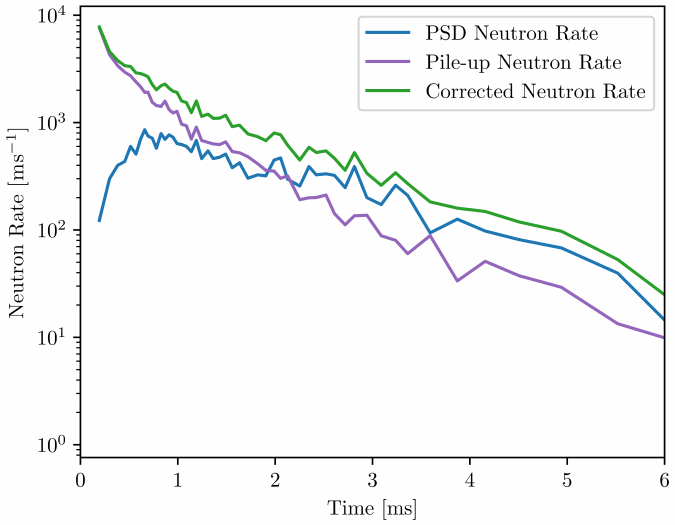}
    \caption{Aggregate neutron rates for shot 22714. Traces correspond to the PSD-identified neutron rate, the rate of pile-up events that are neutrons, and the corrected PSD neutron rate accounting for pile-up events.}
    \label{fig:psd-example-bins}
\end{figure}

Using this approach, the size of a bin is the time difference between subsequent bin start times. The PSD neutron rate for a bin is the number of neutron events in a bin divided by the time-varying bin size and is defined as occurring at the average neutron event time in the bin. The aggregate PSD neutron rate for an example shot is shown in Fig.~\ref{fig:psd-example-bins}. 

The calculation of the rate uncertainty also follows \cite{Howard_PCS16}. A one-standard-deviation uncertainty of the true value of the Poisson rate parameter, given an observed number of events, N, is:
\begin{equation}
\lambda \pm \sqrt{\lambda} = N.
\end{equation}
The solutions $\lambda$ can be found by solving a quadratic equation. For solutions $\lambda = {\{\lambda_1,\lambda_2\}}$, the bounds on the event rate in a bin with time width $\Delta t$ are:
\begin{equation}
\frac{\lambda_1}{\Delta t} < \frac{N}{\Delta t} < \frac{\lambda_2}{\Delta t}.
\end{equation}
For a bin count threshold of 30 events, the solutions are $\lambda = {\{25, 36\}}$. The upper and lower uncertainties on the PSD neutron rate are defined as the respective differences between the upper and lower rate bounds and the measured PSD neutron rate. Note that the upper and lower uncertainties are not symmetric.

Although the general trend of the neutron detection rate from a decaying plasma is to continuously decrease over time, neutron events late in the shot are possible, as shown in Fig.~\ref{fig:psd-example}. Since the time of a binned rate measurement is assigned to be the average of the event time within the bin, if the time to record the threshold number of events for a bin becomes long compared to the decay rate of the plasma emission, then the average rate can significantly exceed the instantaneous rate toward the end of the bin. 
Finally, to obtain the neutron yield and uncertainties at the times for the calculation of the ion temperature, the neutron rates are interpolated in time using linear interpolation in $\log(Y)$ space from the irregular bin measurement times to the desired measurement times at which the density profile is reconstructed.
Due to the decreasing accuracy of this interpolation at very late time, as well as for reasons related to the reliability of the density measurement over time, which are discussed in Section~\ref{plasma-density}, this analysis only includes neutron events occurring in the first $6\mathrm{\,ms}$ of the PI3 plasma discharge.

\subsection{Pile-up Correction}\label{pile-up}
The process to correct for neutron events excluded from PSD due to ``non-hidden'' pulse pile-up starts by estimating the fraction of all events (neutron or gamma ray) that are neutrons. For a single scintillator, taking the total number of neutrons determined by PSD as $N_{\mathrm{PSD,n}}$, and the number of gamma rays determined from PSD as $N_{\mathrm{PSD,\gamma}}$, the probability that an event is a neutron is therefore estimated as:
\begin{equation}
    P_{\mathrm{n}} = \frac{N_\mathrm{PSD,n}}{N_\mathrm{PSD,n}+N_\mathrm{PSD,\gamma} }.
\end{equation}
For simplicity, the probability that an event is a neutron is assumed to be constant throughout the duration of a shot for a given scintillator. As it is expected that this neutron event probability will vary from shot to shot and that scintillators may have different $\mathrm{n}/\gamma$ ratios due to spatial variation in the radiation field, $P_{\mathrm{n}}$ for each scintillator is recalculated for each shot.

The pile-up events excluded from PSD are sorted into the time bins created in Section~\ref{rate-binning}. The time-resolved pile-up event rate, $\dot{N}_{\text{pile-up}}$, is the number of pile-up events in a bin divided by the bin width. The estimated neutron rate from the pile-up events, $\dot{N}_{\text{pile-up,n}}$, is
\begin{equation}
\dot{N}_{\text{pile-up,n}} = P_{\mathrm{n}} \times \dot{N}_{\text{pile-up}} .
\end{equation}
An example of $\dot{N}_{\text{pile-up,n}}$ is shown in Fig.~\ref{fig:psd-example-bins}. The uncertainties of the pile-up neutron rate are calculated in the same way as for the PSD neutron rates. However, in the case of pile-up counts, the number of counts occurring within each time bin varies, so the Poisson rate parameter is recalculated for each time bin. The pile-up neutron rate is added to the PSD neutron rate, $\dot{N}_{\mathrm{PSD,n}}$, to produce a corrected aggregate scintillator neutron rate, $\dot{N}_{\mathrm{n}}$:
\begin{equation}
\dot{N}_{\mathrm{n}} = \dot{N}_{\mathrm{PSD,n}} + \dot{N}_{\text{pile-up,n}}.
\label{eqn:neutron_rate}
\end{equation}
This is shown in Fig.~\ref{fig:psd-example-bins}. Overall, pile-up events will occur at a rate roughly proportional to the true rate, simply due to the PSD tail integration time window being a larger fraction of the inter-arrival time when the rate is high.  The trend of the pile-up rate apparent in Fig.~\ref{fig:psd-example-bins} agrees well with the theoretical rate using the Poisson process probability that two detection events are far enough apart to be distinguishable as separate pulses, yet the latter pulse still occurs within the tail time window of the first pulse.

\subsection{Simulation of Absolute Neutron Detection Efficiency}
\label{neutron-efficiency}
To infer plasma $T_i$ from the neutron count rate, it is first necessary to accurately extrapolate the set of detector counts back to the neutron yield originating from the plasma source. 
Each scintillator has a position-dependent absolute neutron detection efficiency, which serves as a sensitivity calibration on which to base this extrapolation back to the source yield. 
The best available means of determining this set of detection efficiencies is by numerical simulation. 

The UK Atomic Energy Authority (UKAEA) have developed a model of the baseline geometry of PI3 with defined material compositions, and constructed an equivalent, simplified model using the Monte Carlo N-Particle (MCNP) radiation transport code suitable for the analysis of the liquid scintillator response \cite{Valentine_2023}. MCNP6.2 was selected as the Monte Carlo radiation transport code as it offers a suitable combination of charged particle transport and the ability to model complex full-machine geometries \cite{goorley_2012}. This radiation transport model includes the scintillators, all components of the machine, and other significant structures in the facility shown in Fig.~\ref{fig:scint-locs}. The scintillators themselves have been approximated as scintillator material surrounded by shielding, omitting any internal detail of the detectors.

\begin{figure}
    \centering
    \includegraphics[width=0.5\linewidth]{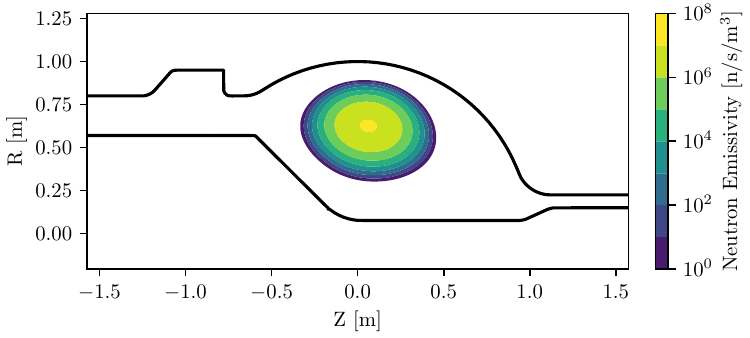}
    \caption{
    Model of PI3 neutron emissivity, used for calibration of detector efficiencies. Derived from a reconstructed plasma state in PI3 shot 18669 at $t=5\mathrm{\,ms}$.}
    \label{fig:neutron-luminosity}
\end{figure}

For the purposes of the calibration calculation, a representative neutron emissivity source with a reasonable spatial distribution was needed.
A Bayesian reconstruction \cite{tancetti_thermal_2025} of a well-studied PI3 discharge (18669) was used at a relevant time of interest ($t=5\,\mathrm{ms}$) along with the assumption of equal electron and ion temperatures (see also \cite{kumar_gyro_submitted}).
A simulated neutron source term was generated from a 2D $(R, Z)$ grid of neutron emissivity (n/s/m$^3$) and ion temperatures (eV) resulting in a volumetric source of isotropic emission (see Fig.~\ref{fig:neutron-luminosity}). Locally isotropic emission is a reasonable assumption for a Maxwellian equilibrium of an Ohmically heated spherical tokamak plasma, however if significant anisotropy is present in the velocity distribution of the plasma ions, such as what arises from neutral beam heating then it would be more accurate to use a more sophisticated source emission profile such as described by~\cite{goncharov_integrated_2024}.
The volumetric source used a D-D neutron energy distribution built into MCNP based on a Gaussian broadened spectrum centered on a $2.45\mathrm{\,MeV}$ peak and broadened by the simulated ion temperature values. A total of $10^9$ source neutron paths were generated in the MCNP simulation used to determine the calibration.

The primary functionality of MCNP includes all the necessary physics, material nuclear properties, and geometry handling capabilities required to model neutron transport.
An approach was developed by UKAEA to utilize the particle track output (PTRAC) from MCNP to record event-by-event data, recording relevant interactions in the scintillators, such as secondary proton and electron energy depositions due to incident neutrons and photons, respectively.
Post-processing is then able to implement tallies with specified energy thresholds, while also applying known detector response curves to simulate scintillator equivalent pulse heights for each neutron interaction \cite{Eljen}. This approach was also previously used in the method of \cite{Howard_PCS16}.

For the calibration simulations, MCNP was configured to record detailed, event-by-event particle interaction data within the scintillator volumes using the PTRAC feature. PTRAC output was enabled with a cell filter restricting recorded events to the scintillator cell IDs only, reducing output size while retaining all relevant detector interactions. Maximum verbosity was selected (WRITE=ALL), with per-event recording of particle position, direction, kinetic energy, time, interaction type, and all banked secondary particles.

No global low-energy cutoffs were imposed to emulate detector discriminator thresholds within MCNP itself. Instead, energy thresholds corresponding to detector response and electronics were applied during post-processing of the PTRAC database. To ensure complete capture of low-energy energy deposition processes relevant to scintillator response, explicit transport cutoffs were set using MCNP CUT cards at 1 keV for photons, electrons, and protons, and zero cutoff for neutrons.

Neutron, photon, electron, and proton transport were enabled using MODE n p e h, allowing consistent treatment of neutron-induced charged-particle production and subsequent energy deposition within the scintillator material. Recoil proton production from neutron interactions was explicitly enabled using the PHYS:N 6J 1 card to ensure accurate modeling of elastic scattering processes that dominate scintillator neutron response.

For both neutron and photon interactions within a scintillator, their detection is always indirect via a secondary charged particle: proton and electron, respectively. In the case of electrons, the energy deposition is directly proportional to the signal amplitude, so a linear energy calibration is applied. However, for protons, there is a non-linear relationship between the initial proton energy and the signal amplitude of the light pulse \cite{verbinski_1968}. 
This nonlinearity needs to be addressed to accurately determine an overall calibration between the source neutron yield and the count of scintillation detection events above a given threshold energy. 

\begin{figure}
    \centering
    \includegraphics[width=0.5\linewidth, trim={0 2cm 0 0}, clip]{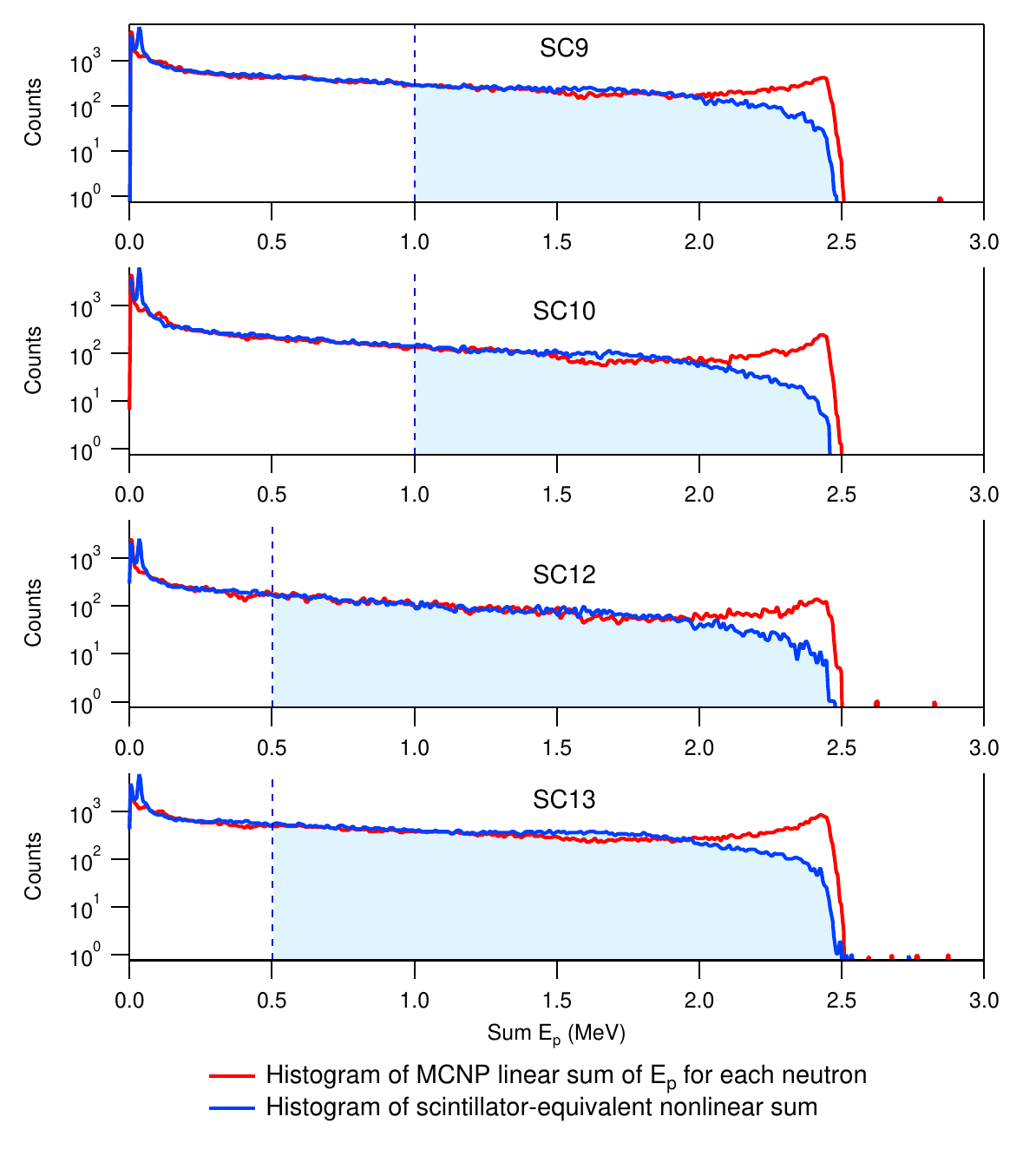}
    \caption{MCNP simulation output data of scintillator proton recoil histograms showing the distribution of both the linear sum of proton energies for each incident neutron (red), and the nonlinear sum (blue) which is equivalent to the actual scintillator response to the neutron flux. The light blue area is the integral above the PSD threshold energy (dashed) which determines the calibration of absolute detection efficiency for each scintillator in its given position. The overall effect of correctly summing the constituent proton energies is to reduce this calibration factor compared to the simple linear sum.}
    \label{fig:MCNP_scint_hist}
\end{figure}

The key to tabulating histograms (see Fig.~\ref{fig:MCNP_scint_hist}) using MCNP in a manner that correctly corresponds to actual scintillator performance, is the fact that a single high-energy neutron entering the detector is likely to collide with several protons, each of which generates a pulse of light. The scintillator PMT will see the sum of the nearly coincident set of light pulses from possibly many ejected protons, but because of the nonlinear relation between the proton's initial energy and the light output, this will not be equivalent to the light output of the sum of all the proton energies (red curves in Fig.~\ref{fig:MCNP_scint_hist}).  
Instead, each proton recoil energy, $E_p$, is converted to an electron-equivalent recoil energy, $E_{ee}$: 
\begin{equation}
    E_{ee}=AE_p-B(1-e^{-CE_p^D}),
    \label{eqn:recoil_conversion}
\end{equation}
where $A$ = 0.83, $B$ = 2.82, $C$ = 0.25, $D$ = 0.93 \cite{cecil_1979, katz_1972}.
The electron-equivalent recoil energies are summed and this value is subsequently converted back into a proton recoil energy. This final inversion of Eq.~\eqref{eqn:recoil_conversion} is done numerically, giving the blue curves shown in Fig.~\ref{fig:MCNP_scint_hist}.

From the resulting histograms of tabulated $E_{p}$ values for each scintillator which depend on the detector's specific location within the modelled neutron field, the absolute detection efficiency is defined as the total histogram counts above the chosen energy threshold divided by the total number of neutrons emitted by the source region in the simulation. 
As such, these absolute efficiency calibration factors $\mathcal{E}_i(E_\mathrm{thr})$ depend on the choice of the minimum energy to include in the histogram count. Low energy events are excluded from this calibration due to lack of clear $\mathrm{n}/\gamma$ discrimination of the PSD method at very low energy.

A stochastic approach has been implemented to calculate the standard deviation on the simulated efficiency values, which does not account for physical uncertainties. The calculated absolute neutron detection efficiencies $\mathcal{E}_i$ and their uncertainties at the corresponding proton recoil energy threshold, $E_\mathrm{thr}$ for each scintillator are listed in Table~\ref{tab:scint-sum}.

\subsection{Conversion to Neutron Yield}
\label{neutron-yield}

For each scintillator, $i$, the neutron count rate, $\dot{N}_i$, can be converted to an estimate of the plasma neutron yield, $Y_i$, in neutrons/second using the absolute neutron detection efficiency for the selected proton recoil energy threshold, $\mathcal{E}_i(E_\mathrm{thr})$: 
\begin{equation}
    Y_i= \frac{\dot{N}_i(E\ge E_\mathrm{thr})}{\mathcal{E}_i(E_\mathrm{thr})}  
    \label{eqn:scint_yield}.
\end{equation} 

\begin{figure}
    \centering
    \includegraphics[width=0.5\linewidth]{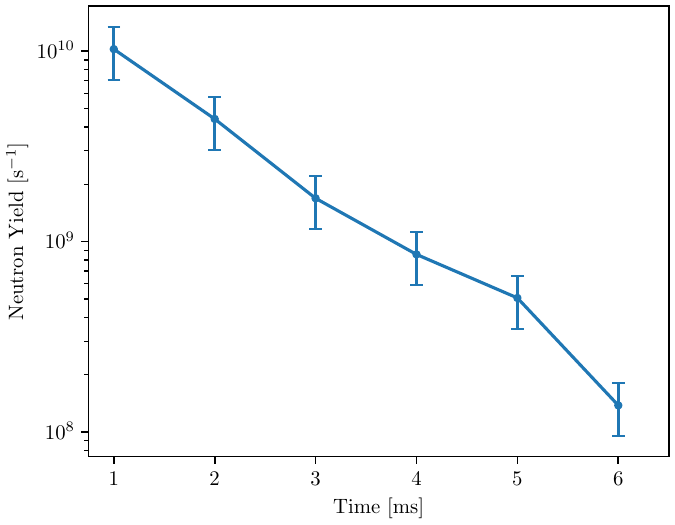}
    \caption{ Neutron yield (s$^{-1}$) for shot 22714 interpolated to ion temperature calculation times.}
    \label{fig:neutron-yield-fig}
\end{figure}

The estimates of source yield $Y_i$ will vary between scintillators due to scattering effects that are not perfectly modeled by the MCNP simulation, and to statistical fluctuations in the count rate intrinsic to the random nature of the trajectories of scattered neutrons as they are transported outward from the plasma. To combine this set of measurements into a single more accurate average estimate of the yield, the method developed in \cite{Howard_PCS16} was followed, where $\mathcal{E}_i(E_\mathrm{thr})$ acts as a weighting factor, giving more weight to scintillators that are more likely to detect a neutron from the source. The resulting neutron yield from the plasma then simplifies to:
\begin{equation}
   Y = \frac{\Sigma_i \dot{N}_i(E\ge E_\mathrm{thr})}{\Sigma_i \mathcal{E}_i(E_\mathrm{thr})} = \frac{\dot{N}_\mathrm{n}}{\mathcal{E}_\mathrm{tot}}
    \label{eqn:avg_total_yield},
\end{equation}
where $\dot{N}_\mathrm{n}$ is the pile-up-corrected aggregated scintillator neutron rate determined in Eq.~\eqref{eqn:neutron_rate} and $\mathcal{E}_\mathrm{tot}$ is the sum of the MCNP-derived scintillator absolute efficiencies. An example plasma neutron yield is shown in Fig.~\ref{fig:neutron-yield-fig} for shot 22714 which had a total time-integrated yield of $2.5 \times 10^7$ neutrons emitted from the plasma.

\subsection{Plasma Density and Volume}\label{plasma-density}
\begin{figure}
     \centering
     \includegraphics[width=0.5\linewidth]{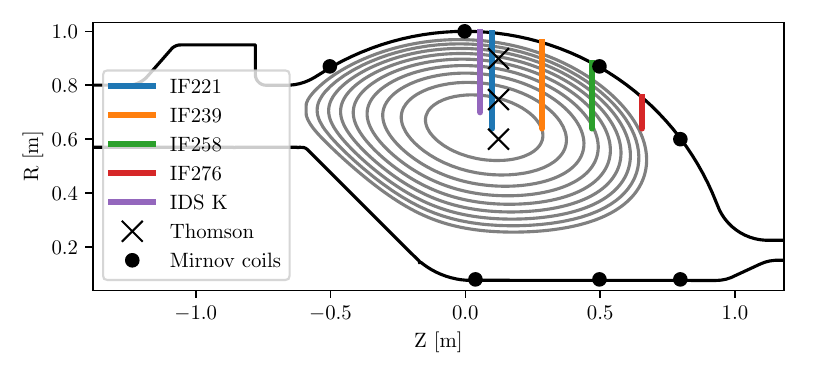}
     \caption{PI3 diagnostic measurement locations projected on the (R, Z) plane relative to the  vessel wall (black) for the interferometer (IF), Mirnov coil, ion Doppler (IDS) and Thomson scattering systems. Poloidal flux contours (grey) for an example equilibrium are shown with the plasma core at a radius of $\sim 600 \mathrm{\,mm}$. IDS and IF chords are projected onto the poloidal plane to show radial extent of measurement.}
     \label{fig:diags-xsec}
 \end{figure}

\begin{figure}
     \centering
     \includegraphics[width=0.5\linewidth]{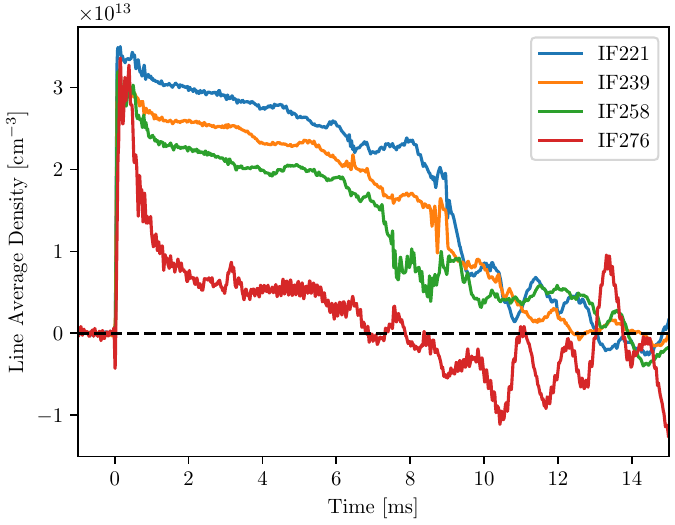}
     \caption{Line average density interferometer data from shot 22714.}
     \label{fig:interf-data}
 \end{figure}

A Bayesian reconstruction algorithm is used to determine the distribution of likely magnetic equilibria and density profiles that best fit Mirnov coil magnetic field measurements and interferometer line average electron density measurements \cite{tancetti_thermal_2025, BayesianRecon_APS2022}. The reconstruction algorithm assumes density is a flux function. Positions of Mirnov coils and interferometer chords are shown in Fig.~\ref{fig:diags-xsec}. Line average electron densities for an example shot are shown in Fig.~\ref{fig:interf-data}. For this analysis, a dilution factor of $1$ is assumed and thus electron and deuterium ion densities are equal. This gives a lower bound on the ion temperature because, for the same yield, a lower ion density requires a higher ion temperature.

\begin{figure}
    \centering
    \includegraphics[width=0.5\linewidth]{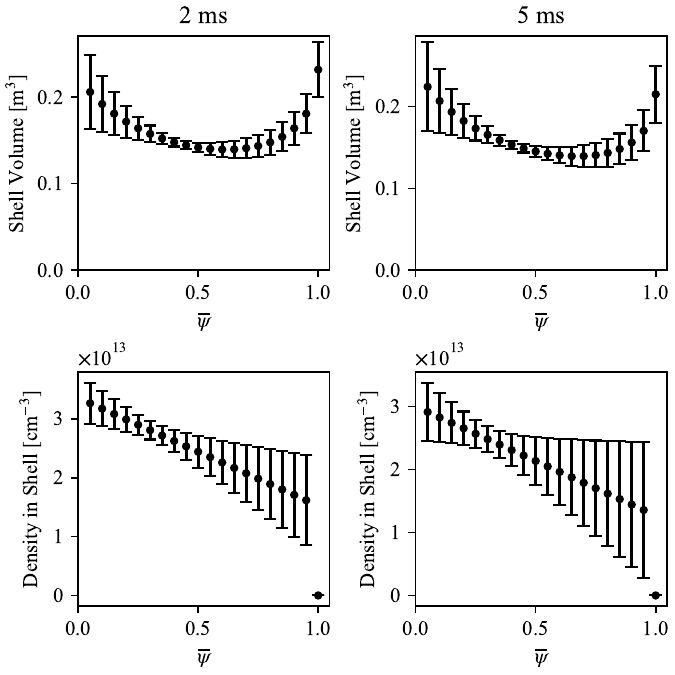}
    \caption{Example reconstructed density profile with uncertainties for shot 22714 at $2~\mathrm{ms}$ and $5~\mathrm{ms}$. There are 20 shells evenly spaced in normalized poloidal flux from $\bar{\psi}=0$ (core) to $\bar{\psi}=1$ (last closed flux surface). Density at the last closed flux surface is pinned to zero.}
    \label{fig:density_volume}
\end{figure}

The reconstruction algorithm outputs the density and volumes for 20 poloidal flux shells (Fig.~\ref{fig:density_volume}) evenly spaced in poloidal flux from the core to the last closed flux surface (LCFS) . The interferometer density measurement is susceptible to vibration error which become significant after $t = 7 \mathrm{\,ms}$ (Fig.~\ref{fig:interf-data}). This can drive low-density measurements at the edge-most chords negative, resulting in unreliable density profiles after this time. Analysis therefore uses data up to $6\mathrm{\,ms}$, for which density profiles are reliable.

Magnetic reconstruction uses an axisymmetric equilibrium model of the plasma. This is considered a reasonable model since diagnostics are distributed at several toroidal angles and show good agreement with reconstructed axisymmetric synthetic measurements.

\subsection{Temperature Profile}\label{temp-profile}

The ion temperature profile is parametrized as:
\begin{equation}
T_i(\overline{\psi}) = T_i(0)\left[1-\overline{\psi}^2\right]^\alpha
 \label{eqn:Ti_prof},
\end{equation}
where $\overline{\psi}$ is the normalized poloidal flux with $\overline{\psi} = 0$ at the magnetic axis (core) and $\overline{\psi} = 1$ at the last closed flux surface. The parameter $\alpha$ defines the profile’s peakedness. Two $\alpha$ values are considered in this work to study the sensitivity of the inferred core temperature to profile shape: $\alpha=0$, corresponding to a flat profile; and $\alpha=1$, corresponding to a parabolic peaked profile.

\begin{figure}
    \centering
    \includegraphics[width=0.5\linewidth]{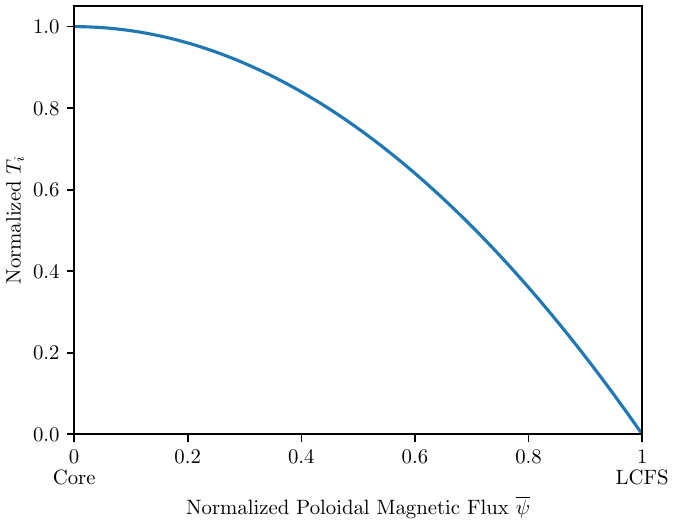}
    \caption{Parabolic ion temperature profile model ($\alpha=1$).}
    \label{fig:ti-profile-model}
\end{figure}

\begin{figure}
    \centering
    \includegraphics[width=0.5\linewidth]{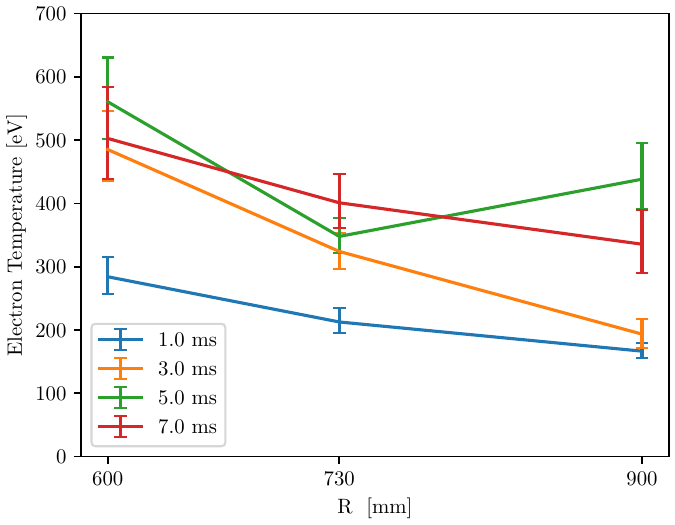}
    \caption{Multi-time Thomson scattering electron temperature profile for shot 22714 showing a peaked profile. R=600~mm is the core measurement, R=900~mm is the edge-most measurement. The last closed flux surface is around R=1000~mm.}
    \label{fig:ti-profile-22714}
\end{figure}

An example of a parabolic peaked temperature profile is shown in Fig.~\ref{fig:ti-profile-model}. The ion temperature profile of PI3 plasmas is not constrained by measurement, although the most likely temperature profile is believed to be peaked based on Thomson scattering electron temperature profile measurements \cite{PI3_APS2024}, with an example shown in Fig.~\ref{fig:ti-profile-22714}.

These two temperature profiles provide a way to gauge certain systematic uncertainties in the ion temperature. 
The parabolic profile is a reasonable but optimistic interpretation of a population of more energetic ions concentrated along the core reconstructed magnetic flux surfaces, while the flat profile is the logical limit of a plasma with high thermal transport or a profile effectively broadened by large orbits of energetic ions.

\section{Ion Temperature Calculations}\label{temp-calcs}

Assuming a Maxwellian distribution for ion velocity (section~\ref{sec:maxwellian_assumption}), the ion temperature can be determined from the measured neutron yield using an interpolated inverse lookup of a pre-tabulated forward-model built using the measured density profile as an input. 
For each of the poloidal flux shells, the neutron yield is:
\begin{equation}\label{eqn:shell_yield}
        Y_\mathrm{shell} = \frac{1}{2} \langle \sigma v \rangle_\mathrm{shell}n_{d,\mathrm{shell}}^2 V_\mathrm{shell},
\end{equation}
where $\langle \sigma v \rangle_\mathrm{shell}$, $n_{d,\mathrm{shell}}$ and $V_\mathrm{shell}$ are respectively a shell's reactivity, deuterium ion density and volume.\footnote{This formula for yield differs by a  factor two from that used in \cite{Howard_PCS16} because in that reference the total D-D fusion reactivity was used, working from \cite{NRL}, then assuming a branching ratio 1/2 for neutron production. The present work uses the reactivity formula from \cite{bosch_improved_1992} for the specific reaction $\mathrm{ D(d,n){}^3He}$ that produces a neutron. Both approaches yield equivalent results.}
The reactivity $\langle \sigma v \rangle_\mathrm{shell}$ for the $\mathrm{ D(d,n){}^3He}$ reaction is a fast function of local ion temperature for which a Bosch-Hale formula is used \cite{bosch_improved_1992}.
It is through this strong functional dependence that it becomes possible to estimate $T_i$ with reasonable accuracy from the plasma neutron yield.

The overall neutron yield from the whole plasma, $Y$, is given by the sum over shells:
\begin{equation}\label{eqn:total_yield}
    Y = \frac{1}{2}\int \langle \sigma v \rangle n_{d}^2\,dV \approx \sum Y_\mathrm{shell}.
\end{equation}
To relate the core ion temperature $T_i$ to $Y$, the average reactivity, $\overline{\langle\sigma v\rangle}$, is introduced:
\begin{equation}
    \overline{\langle\sigma v\rangle}\equiv \frac{Y}{\!\int\!\frac12n_d^2\,dV}.
\end{equation} In a forward-model, for a given temperature profile shape and density profile, the average reactivity is a function of $T_i$.  A lookup table of this function is calculated (Section~\ref{rate-reactivity}) and then interpolation is used to obtain both $T_i$ and its error bars from experimental measurements of the neutron yield (Section~\ref{Ti_calc}).

\subsection{Construction of Average Reactivity Lookup Table}\label{rate-reactivity}

The calculation begins by defining the ion temperature profile functional form as parameterized by Eq.~\eqref{eqn:Ti_prof} for a particular choice of $\alpha$, and then specifying the full range of core ion temperature values $(T_{i,\mathrm{min}}, T_{i,\mathrm{max}})$ to include in a forward-model table,  which spans the range of physically plausible values over the duration of a shot. The range of temperatures in this set limits the temperatures that can be predicted. Each core temperature is used to scale the profile shape, yielding a set of ion temperature profiles. For each temperature profile, the shell reactivities, $\langle \sigma v \rangle_\mathrm{shell}$, are calculated from the shell temperature using a Bosch-Hale formula \cite{bosch_improved_1992}.

Using the nominal (reconstruction mean) shell density and volume from the experiment,
summing over the shell neutron yields following Eq.~\eqref{eqn:shell_yield} produces the volume-integrated neutron yield, $Y_{T_{i,\mathrm{test}}}$, from Eq.~\eqref{eqn:total_yield}.

The average test reactivity, $\overline{\langle\sigma v\rangle}_{T_{i,\mathrm{test}}}$, is then tabulated from $Y_{T_{i,\mathrm{test}}}$ as follows:
\begin{equation}\label{avg-reactivity-eq}
    \overline{\langle\sigma v\rangle}_{T_{i,\mathrm{test}}} = \frac{2Y_{T_{i,\mathrm{test}}}}{\sum n_{d,\mathrm{shell}}^2 V_\mathrm{shell}}.
\end{equation} 
This forward-model table stores the average reactivity for each temperature value.
It is worth noting that the average test reactivity is independent of an overall density scaling, depending only on the shape of the density profile. This table is recomputed for each timestep of reconstructed density profile data, but then held constant for each iteration of the Monte Carlo error analysis described in Section~\ref{Ti_calc}.

\subsection{Ion Temperature Calculation}\label{Ti_calc}

The ion temperature can be determined by interpolating in the tabulated forward-model average test reactivity table to match the average reactivity calculated directly from the observed neutron yield and reconstructed density and volume profiles.

For error analysis, this basic interpolation algorithm is wrapped in a Monte Carlo (MC) framework to calculate both the ion temperature and its overall measurement uncertainty. The input parameters of the overall $T_i$ calculation are the shell density,  shell volume, scintillator-detected neutron rate, $\dot{N}_\mathrm{n}$, and total absolute neutron detection efficiency, $\mathcal{E}_\mathrm{tot}$, in addition to uncertainty values for each of those inputs.  The calculation involves a large number of MC runs. Each MC run uses values of the input parameters independently obtained by randomly sampling a normal distribution with mean $\mu$ given by the measurement value and $\sigma$ given by the uncertainty of that parameter.  These sampled values are denoted by the subscript $\mathrm{MC}$.  Each run gives an average MC reactivity, denoted $\overline{\langle\sigma v\rangle}_\mathrm{MC}$, that is calculated as
\begin{equation}\label{reactivity_interp}
    \overline{\langle \sigma v \rangle}_\mathrm{MC} = \frac{2Y_\mathrm{MC} }{\sum n_{d,\mathrm{shell,MC}}^2 V_\mathrm{shell,MC}},
\end{equation} 
where:
\begin{equation} \label{eq:Ymc}
Y_\mathrm{MC} = \frac{\dot{N}_\mathrm{n,MC}}{\mathcal{E}_\mathrm{tot,MC}},
\end{equation}
following Eq.~\eqref{eqn:avg_total_yield}.
The core ion temperature is estimated by interpolating the average MC reactivity, $\overline{\langle\sigma v\rangle}_\mathrm{MC}$, from the set of average test reactivities, $\overline{\langle\sigma v\rangle}_{T_{i,\mathrm{test}}}$, computed in Section~\ref{rate-reactivity}.  

The result of all the MC runs is an ensemble of core $T_i$ values. The experimentally determined $T_i$ is taken as the 50th percentile value, and the upper and lower uncertainties are defined as the 16th and 84th percentile values. This process results in a core ion temperature with uncertainty for a given temperature profile shape. 

A convergence analysis was performed to determine the minimum number of MC runs required for the variance in the ion temperature to converge. It was found that \mbox{$500$} runs were sufficient for the change in the uncertainty estimate after each MC run to fall below \mbox{$10\mathrm{\,eV}$}. All figures in Sections~\ref{temp-calcs} and \ref{results} use 10000 MC runs, giving an incremental ion temperature uncertainty change of less than \mbox{$3\mathrm{\,eV}$}.

\section{Experimental Results}\label{results}

\begin{figure}
     \centering
     \includegraphics[width=0.5\linewidth]{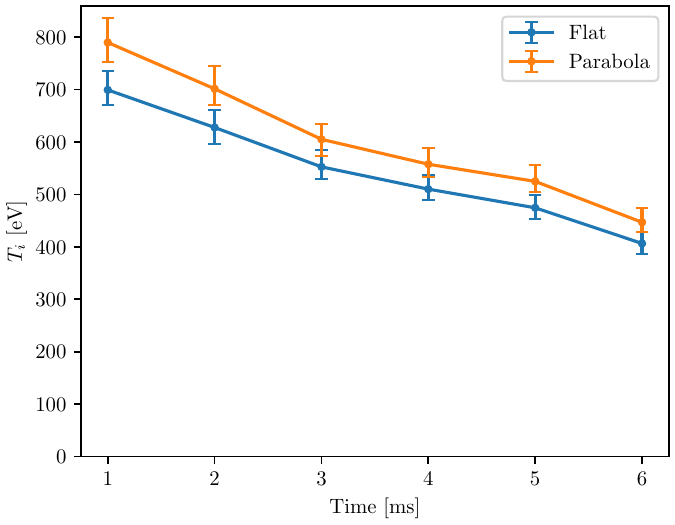}
     \caption{$T_i$ calculated from neutron scintillator counts for PI3 shot 22714 for flat and parabolic $T_i$ profiles. Mean values are represented as the 50th percentile and upper and lower limits are represented by the 16th and 84th percentiles.}
     \label{fig:22714_Ti}
\end{figure}

\begin{figure} 
     \centering
     \includegraphics[width=0.5\linewidth]{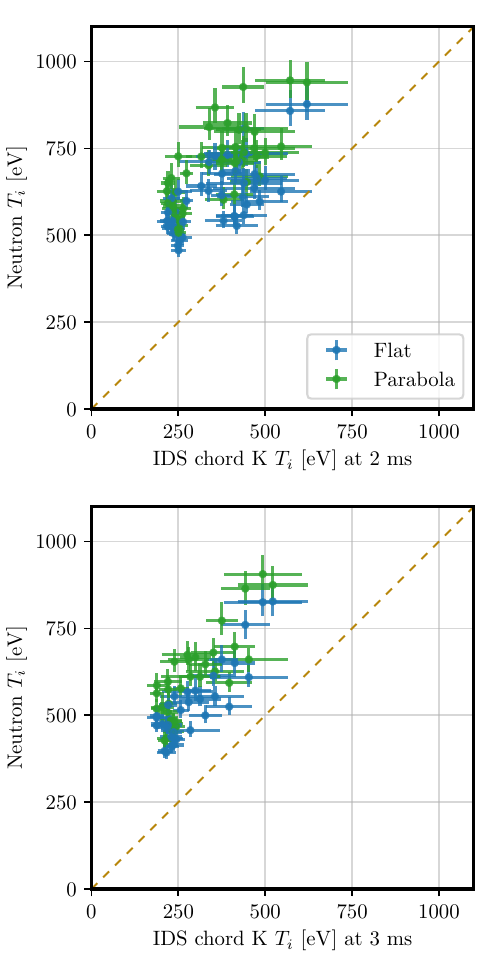}
     \caption{ Neutron counting $T_i$ plotted against IDS $T_i$ from a single chord for 33 discharges in PI3. {\sl Top:} at $2\mathrm{\,ms}$ into the discharge {\sl Bottom:} at $3\mathrm{\,ms}$ into the discharge. Values using both flat and parabolic density profile models are plotted. The IDS $T_i$ values are generally smaller than the neutron counting derived values.
     }
     \label{fig:IDSvsneutron}
\end{figure}

Neutron yield and interferometry data were processed to calculate $T_i(t)$ for 207 plasma shots from PI3 performed between March and August 2024. Fig.~\ref{fig:22714_Ti} shows the resulting time-resolved $T_i(t)$ for PI3 shot 22714 considering the cases of flat and parabolic temperature profile shapes. 
The main source of uncertainty in this measurement is the unknown ion temperature profile shape, different values for which may result in moderately distinct values of $T_i$ that deviate more than the intrinsic error given a particular profile. 
Therefore, when using these neutron-inferred $T_i(t)$ data (such as to calculate thermal energy confinement time), the wider error bars determined from the full range of values are used on the set of plausible profiles.  

A comparison of ion temperatures derived from (a) neutron rates using a parabolic profile and (b) a single Ion Doppler Spectroscopy (IDS) chord, for 33 PI3 discharges in the shot range 22331 to~22721 is shown in Fig.~\ref{fig:IDSvsneutron}. The data indicate a general trend of correlation between these two ion temperature diagnostics. 
In contrast to the neutron yield $T_i$ diagnostic, the IDS chord~K does not penetrate fully to the magnetic axis of the plasma, and the C~V charge state being observed (via its spectral line at $227\mathrm{\,nm}$) is likely to not exist as an equilibrium charge state in the high-temperature region of the PI3 core plasma. Instead, maximum emission of that line would be expected significantly off axis, where it is possible for the ionization state to exist in coronal equilibrium (see discussion in \cite{tancetti_thermal_2025, Howard_PCS16}). Thus the IDS $T_i$ values, more representative of the regions outside the core, are expected to be smaller than the neutron $T_i$ values for the same plasma. In addition, one expects the general trend that a plasma with a higher core temperature would also have a higher temperature outside the core. Taken together, the IDS and neutron data are most consistent with a centrally peaked $T_i$ profile similar to the parabolic form.

\section{Assumptions and Limitations}
\subsection{The Assumption of Maxwellian Ions} \label{sec:maxwellian_assumption}

In PI3 experiments, the ion collision time is not much shorter than the time scales of the plasma discharge and, as a result, a fully Maxwellian ion distribution is not guaranteed. 
For example, the ion collision time in a deuterium plasma with $T_i=1\mathrm{\,keV}$ and $n_i=5\times10^{19}\mathrm{\,m^{-3}}$ is $\tau_i=1\mathrm{\,ms}$ \cite{wesson_tokamaks_2011}.
If the formation of the plasma target results in a supra-thermal ion distribution, with more high-energy ions than a Maxwellian, then the rate of fusion reactions will be enhanced for some considerable time. 
In this case, an ion temperature estimated from a neutron rate which assumes a Maxwellian distribution will be higher than the actual kinetic temperature.
Further, the evolution of the neutron rate will be affected by the process of ion thermalization: ion collisions will make the distribution more Maxwellian, causing a decay in the neutron rate in addition to the decay due to ion cooling by transport and collisions with electrons.

It is thus important to highlight that the assumption of a Maxwellian ion distribution is a working assumption here.
The possibility of a supra-thermal ion distribution should not be discounted when analyzing MTF experiments with a low density target plasma such as created in PI3.

Future net gain MTF experiments on millisecond time scales use initial plasma targets of higher density, for example $n_i=2\times10^{20}\mathrm{\,m^{-3}}$  \cite{laberge_magnetized_2019}, a density already realized in the PCS-16 experiment \cite{Howard_PCS16}.  In such MTF experiments the density-time product is significantly higher than for PI3 and the assumption of Maxwellian ions is justified.

To verify the Maxwellian assumption, the ion temperature and ion energy distribution can be measured with other diagnostic techniques like neutron time-of-flight spectroscopy \cite{firk_neutron_1979} and a neutral particle analyzer \cite{mezonlin_2007}. A neutron spectrometer is under development for use with highly compressed plasmas in future General Fusion experiments \cite{carle_neutron_2022}.

\subsection{Neutron Count Rate Limitations}
PI3 shots typically provided sufficient neutrons ($>$30 per time bin) that low neutron counts ($<$5 per time bin) were not a challenge. To estimate a lower bound on temperatures this diagnostic can measure, combining equations \eqref{eqn:avg_total_yield} and \eqref{eqn:shell_yield}, and assuming a constant density profile gives:
\begin{equation}
    \label{eqn:loweryield}
    \overline{Y}_{\mathrm{min}}(T_{i,{\mathrm{min}}})=\frac{C_{\mathrm{min}}}{\mathcal{E}_{\mathrm{in}}} \frac{4\pi d^2}{N_s A}=\frac{1}{2}  n_{d}^2 V \tau \langle \sigma v \rangle ( T_{i,{\mathrm{min}}}) ,
\end{equation}
where  $\overline{Y}_{\mathrm{min}}$ is a minimum time-integrated neutron yield (counts) required to allow an ion temperature measurement with reasonable error bars that contains useful information about the evolution of the ion temperature as a function of time, for a single plasma discharge. If scattering is negligible, the time-integrated yield is related to the total counts  $C_{\mathrm{min}}$ detected by the scintillators, via the first equality in \eqref{eqn:loweryield}.  In order to make a useful measurement a practical limit at $C_{\mathrm{min}} = 50$ is set, which would allow five neutrons counts in each time bin and 10 time bins measuring the $T_i$ evolution during the plasma duration. To further describe \eqref{eqn:loweryield}, $d$ is the distance from the detectors to the plasma, $\mathcal{E}_{\mathrm{in}} = 0.612$ is the intrinsic neutron detection efficiency (typical value for these scintillators),  $N_s$ is the number of scintillator detectors (four in the case of PI3), $A$ is the area of the scintillators perpendicular to the neutron flux (225~cm$^2$ for SC12 and SC13, 83.6~cm$^2$ for SC9 and SC10), while $V$ is the volume of the plasma,  $\tau$ is the plasma burn time over which the time bins are divided, and $T_{i,\mathrm{min}}$ will be the minimum measurable ion temperature to be determined. Using the Bosch-Hale formula to relate reactivity to ion temperature, and setting the detectors at an average distance of 5 meters yields a prediction of 350~eV as the minimum measurable temperature for the PI3 configuration having $n_d =10^{19}~$m$^{-3}$ to achieve a $T_i$ measurment with 10 time bins over $\tau  = 20$ ms, with each bin having at least $N>5$ counts per bin. At this lower bound case, the $T_i$ measurement would have a relative uncertainty of 50\%, mostly due to intrinsic error in the density measurement. In general, the method could be improved by adding more detectors and putting them closer to the machine, and in the case where 100 detectors are set at a distance of 2~meters, this method could measure down to 200~eV. In the case where the plasma is even colder, combining detector counts across similar plasma shots could be done to make an ensemble-average temperature measurement of lower temperatures, provided that neutron count rate from fusion is still significantly above the ambient background rate from solar, cosmic and other natural sources.

There is also an upper bound to the instantaneous rate of neutron flux incident on the detectors that still produces an accurate signal. In Sec.~\ref{pile-up}, the pile-up correction algorithm is described; it accounts for events that begin before the preceding pulse has fully returned to baseline. When a pile-up event occurs within 10~ns of the previous peak, the Python find$\_$peaks algorithm will fail to identify a new peak, undercounting the pile-up and/or potentially falsely flagging two gamma events as one-neutron event if the second peak has the effect of broadening the tail of the identified peak. The frequency of these hidden and problematic events can be estimated using the highest event rate. 

Assuming that events follow a Poisson time-arrival distribution, the probability that two events occur within a time interval 
$t_0$ is given by: 
\begin{equation}
    P(\Delta t<t_0)=1-e^{-\lambda t_0},
\end{equation}
where $\lambda$ is the local event rate. In the case where the maximum $\lambda$ is greater than 6.7$\times$10$^7$ counts/s, this effect would introduce a local error in count rate of 50~\%. For extremely high count rates exceeding $10^9$~counts/s the pile-up effect can fully saturate where all pulses overlap and if this continues for long enough it will have the effect that the gain of the PMT detector will significantly reduce due to dynode voltage droop, which can take 0.1 - 0.5~ms to recover as the dynode capacitors recharge. Bench characterization of the scintillator's PMTs show that they begin to experience gain droop when the time integral of the output signal is larger than $10^{-6}$~Volt-seconds within a time of 10~$\mu$s. Overall, the case of count rate being too high can usually be improved by having the detectors positioned farther away from the source, as the count rate would diminish as the square of the distance from the scintillator to the plasma. Ideally, design of an experimental arrangement could be optimized with this analysis given an expected range of temperatures and densities for the plasma.

\section{Conclusion}\label{conclusion}
This paper presents a method for inferring deuterium ion temperature ($T_i$) from neutron counts measured with fast liquid scintillators, developed with data from General Fusion's Plasma Injector~3 (PI3). The method allows for the determination of $T_i$ without a direct line-of-sight to the plasma or neutron collimation.

Complementing this technique with a direct measure of the velocity distribution of the deuterium ions \cite{carle_neutron_2022} will be valuable to further justify the assumption of a Maxwellian distribution.

A multi-step computational process calculates the neutron yield through PSD of scintillator events. 
MCNP modeling of scintillators in the PI3 lab space provides absolute neutron detection efficiencies that account for neutron scattering through the machine and facility components. 

The temperature calculation uses the measured neutron yield, as well as the density profile and geometry of the flux surfaces of the plasma as inputs. 
The density profile and flux surface geometry are calculated using a magnetic reconstruction algorithm based on measured diagnostic information on the plasma magnetic fields and the line-average plasma density. 
With the assumption of a $T_i$ profile shape, a set of central ion temperature values is iterated over, and the corresponding neutron yield is calculated using a reconstructed density profile from interferometer measurements. The process converges on the ion temperature value that best matches the experimentally determined neutron yield. This algorithm can be generalized within a Monte Carlo structure to directly estimate the statistical error of the $T_i$ output, given measured and estimated errors of the input quantities. Two temperature profile models are considered as upper and lower bounding cases, and the calculated core temperatures are compared.

Compared to IDS measurements of ion temperature in PI3 plasmas, values of $T_i$ derived from the neutron counting method presented here show the physically expected relationship between the two methods, in that the neutron counting derived measurement is more representative of the hotter plasma core.

Although neutron counting was already being used to infer ion temperature in the early 1970's \cite{artsimovich_investigation_1972}, the continued development of this method remains highly relevant for current technology, as direct diagnostic access to the plasma must decrease as devices transition from plasma experiments into viable power plants.
Specifically in the instance of General Fusion's Magnetized Target Fusion machines, neutron counting can be used to provide a time-resolved measurement of the compressive heating of the plasma to fusion conditions. 

Scintillator neutron-counting diagnostics are well suited to current-generation machines such as PI3, where neutron rates remain moderate. In principle, this diagnostic can continue to be used in future MTF experiments reaching highly compressed plasma states. Accurate anticipation of high neutron rates from these experiments will be required to design the diagnostic arrangement to enable useful $T_i$ measurements. As neutron production scales exponentially with ion temperature, liquid scintillator counters risk saturation, requiring placement at greater distances from the target. This, in turn, increases reliance on neutron transport modeling (e.g., MCNP) to accurately interpret measured signals. Moreover, line-of-sight diagnostics such as interferometry become limited as the plasma is increasingly obscured by the liner during compression, necessitating extrapolation or modeling of the plasma density evolution. While time-of-flight neutron spectrometry will likely become the primary diagnostic for ion temperature in such regimes, scintillator counters can still provide valuable measurements of total neutron yield to inform detector placement and calibration, and offer a second measurement of $T_i$ to validate neutron spectrometer measurements.

\section*{Acknowledgements}
The authors thank Steven C.\ Bradnam, Hari Chohan, Lee W. Packer, and Sean Conroy for their work developing the MCNP models, performing the neutronics simulations based on input data provided by General Fusion, and for advice on the methodology and manuscript. 
The authors thank Prof. Krzysztof Starosta for review of the original manuscript and Alex Woinoski for help with scintillator calibration.
The authors thank Andrea Tancetti for valuable assistance.
This work was supported by funding from the Government of Canada through its Strategic Innovation Fund (Agreement No.~811-811346). The funding source had no involvement in: the study design; the collection, analysis and interpretation of data; the writing of the report; the decision to submit the article for publication. 

\bibliographystyle{ans_js}
\bibliography{refs}

\end{document}